\documentclass[%
 reprint,
 amsmath,amssymb,
 aps,
]{revtex4-2}

\usepackage{graphicx}
\usepackage{bm}
\usepackage{xcolor}
\usepackage{soul}
\usepackage{booktabs}

\DeclareRobustCommand{\edit}[1]{{\color[rgb]{0,0,0}#1}} 
\DeclareRobustCommand{\editS}[1]{{\color[rgb]{0,0,0}#1}} 
\newcounter{appnum}

\begin{document}


\title{Beyond optical chirality density:\\tensor-based description of electromagnetic chirality}


\author{Ilia Smagin}
 \email{Ilia.Smagin@skoltech.ru}
\author{Sergey Dyakov}%
 \email{S.Dyakov@skoltech.ru} 
\affiliation{Skolkovo Institute of Science and Technology, Moscow, Russia}

\date{\today}

\begin{abstract}

Optical chirality density is widely used as a local scalar measure of the chirality of an electromagnetic field and its interaction with isotropic chiral matter. However, within the dipole approximation, a general bi-anisotropic material response probes direction-dependent field bilinears that cannot, in general, be reduced to a single pseudoscalar. This necessitates the development of a systematic tensorial formulation to describe the coupling between arbitrary fields and general reciprocal bi-anisotropic media. To this end, we extend the Tang--Cohen formalism to a general anisotropic optical-activity pseudotensor, treating its trace and traceless symmetric and antisymmetric parts separately. We demonstrate while the trace part couples to classical optical chirality density, the traceless symmetric part is probed by the five Lipkin densities $Z^{ab0}$, which form part of the Lipkin tensor. On the other hand, for the antisymmetric part of the optical-activity pseudotensor, the coupling term is proportional to the reactive part of the local Poynting vector. Finally, we provide numerical examples of pseudo-chiral and anisotropic chiral electromagnetic fields and study their interaction with different bi-anisotropic media. The developed framework offers a systematic tensorial description of chiral and pseudo-chiral light-matter coupling. Importantly, our analysis identifies which specific field quantities must be enhanced to optimize the magneto-electric coupling in such media.

\end{abstract}

\maketitle

\section{\label{sec:Introduction}Introduction}

Chirality is a manifestation of broken mirror symmetry and plays a central role in electromagnetic phenomena that distinguish left from right. In classical electrodynamics, such effects cannot be characterized by energy or momentum alone and require field quantities with the appropriate transformation properties. \edit{In particular}, a local scalar field quantity governing an isotropic chiral light--matter interaction must be even under time reversal and odd under spatial inversion.

The optical chirality density, $\rho_\chi$, originally contained in Lipkin's zilch tensor~\cite{Lipkin1964}, acquired a direct physical meaning through the work of Tang and Cohen~\cite{Tang2010}. In Ref.~~\cite{Tang2010} it has been shown that $\rho_\chi$ governs the excitation-rate asymmetry of a small isotropic chiral molecule. In the units adopted here, $c=\varepsilon_0=1$, it is defined as
\begin{align}\label{OCD def}
\rho_\chi
=
\frac{1}{2}
\left(
-\bm{E}\cdot\frac{\partial\bm{H}}{\partial t}
+
\bm{H}\cdot\frac{\partial\bm{E}}{\partial t}
\right).
\end{align}
Here $\bm E$ and $\bm H$ are the electric and magnetic fields, respectively. The corresponding optical-chirality flux is
\begin{align}\label{OCF def}
\bm{j}_\chi
=
\frac{1}{2}
\left(
\bm{E}\times\frac{\partial\bm{E}}{\partial t}
+
\bm{H}\times\frac{\partial\bm{H}}{\partial t}
\right).
\end{align}
For source-free fields satisfying Maxwell's equations, these quantities obey the continuity equation
\begin{align}\label{OCD div}
\partial_t\rho_\chi
+
\bm{\nabla}\cdot\bm{j}_\chi
=
0.
\end{align}

While optical chirality density satisfies the fundamental symmetry requirements---namely, it is even under time reversal and odd under spatial inversion---it describes only the isotropic pseudoscalar part of electromagnetic chirality. In bi-anisotropic media or structured electromagnetic fields, the chiral light--matter interaction need not be governed by the optical chirality density alone: different spatial directions and symmetry axes of the system can enter the coupling in inequivalent ways. In this sense, electromagnetic chirality can acquire a direction-dependent, tensorial structure.

This observation suggests that, within the electric-dipole--magnetic-dipole level of chiral light--matter interaction, electromagnetic chirality cannot always be fully characterized by a single pseudoscalar quantity. Tang and Cohen~\cite{Tang2010} emphasized an even broader point: no single measure of electromagnetic chirality can be universal for all possible fields and all possible chiroptical responses, because higher multipoles and more complicated material structures may become relevant.

Here, we restrict our attention to the electric‑dipole--magnetic‑dipole regime and investigate the generalization of the scalar Tang–Cohen factorization to the general reciprocal bi-anisotropic molecular response. We will demonstrate that for the symmetric part of the material magneto-electric tensor, the required tensorial field quantities are present in the Lipkin, or zilch, tensor \cite{Lipkin1964}. Whereas for for the antisymmetric part, the coupling term is proportional to the reactive part of the local Poynting vector. Since the symmetric case is much more often, below we pay special attention to properties of the Lipkin tensor.



The Lipkin tensor organizes local bilinear combinations of the electromagnetic field and its derivatives into a third-rank pseudotensor $Z^{\mu\nu\sigma}$, with $\mu,\nu,\sigma=0,1,2,3$. Its components $Z^{\mu\nu0}$ therefore act as densities of conserved quantities, while $Z^{\mu\nu c}$, $c=1,2,3$, are the corresponding flux components. The properties of these quantities are described in detail in Sec.~\ref{sec:Zilch tensor}; for now its only important that the components $Z^{000}$ and $Z^{0k0}=2j_{\chi,k}$ represent the classical optical chirality density and flux, and in the normalization used here,
\begin{align}
Z^{000}=2\rho_\chi,
\qquad
Z^{0k0}=2j_{\chi,k}.
\end{align}

Lipkin introduced these conserved quantities in 1964 and showed that they were not reducible to energy, momentum, or angular momentum~\cite{Lipkin1964}. Shortly thereafter, their mathematical structure and conservation properties were clarified and generalized~\cite{kibble1965conservation,Morgan1964,Candlin}. These early works did not establish a general observable interpretation of the zilch quantities, and Candlin even concluded that they were unlikely to possess a special physical significance~\cite{Candlin}. Subsequent studies related zilch conservation to the symmetries of Maxwell's equations and to Noether's theorem~\cite{Simulik1989_1,Simlulik1989_2,Cameron1,Cameron2,Philbin2013,Bliokh,aghapour2020zilch,letsios2022continuity}.

An important step in the physical interpretation of the Lipkin densities $Z^{ab0}$ was taken by Yang and Cohen~\cite{YangCohen2011}. They considered monochromatic electromagnetic fields and derived the field bilinears that drive electric-dipole--magnetic-dipole transitions in oriented molecules. They then showed that the six quantities $Z^{ab0}=Z^{ba0}$ are linear combinations of these electric-magnetic field bilinears. Thus, the connection between the Lipkin densities $Z^{ab0}$ and molecular observables was established. In a different context, Smith and Strange subsequently related zilch components to the topology of particular nontrivial vacuum electromagnetic fields~\cite{smith2018lipkin}.

The present work develops the componentwise connection identified by Yang and Cohen into a systematic tensorial framework. We refer the six components $Z^{ab0}$ to as the Lipkin densities; these components are organized as a symmetric rank-two pseudotensor. Under spatial rotations, this block decomposes into an isotropic trace component, fixed by $Z^{000}=2\rho_\chi$, and five symmetric-traceless anisotropic components. We then extend the Tang--Cohen excitation-rate formalism to a general anisotropic electric-dipole--magnetic-dipole response and show how Lipkin densities and symmetric part of the optical activity pseudotensor are coupled. Moreover, we show that the antisymmetric part of the optical activity pseudotensor is sensitive to the reactive part of the complex Poynting vector and gives the reciprocal omega-type contribution.

We use point-group symmetry notation to identify materials with anisotropic magneto-electric response exhibiting non-vanishing coupling with certain Lipkin densities.

Finally, we provide specific examples of electromagnetic fields with the enhanced Lipkin densities and demonstrate how they interact with different anisotropic chiral and pseudochiral media.

\section{\label{sec:Zilch tensor}Zilch tensor}
\begin{figure}[t!]
    \centering
    \includegraphics[width=1\columnwidth]{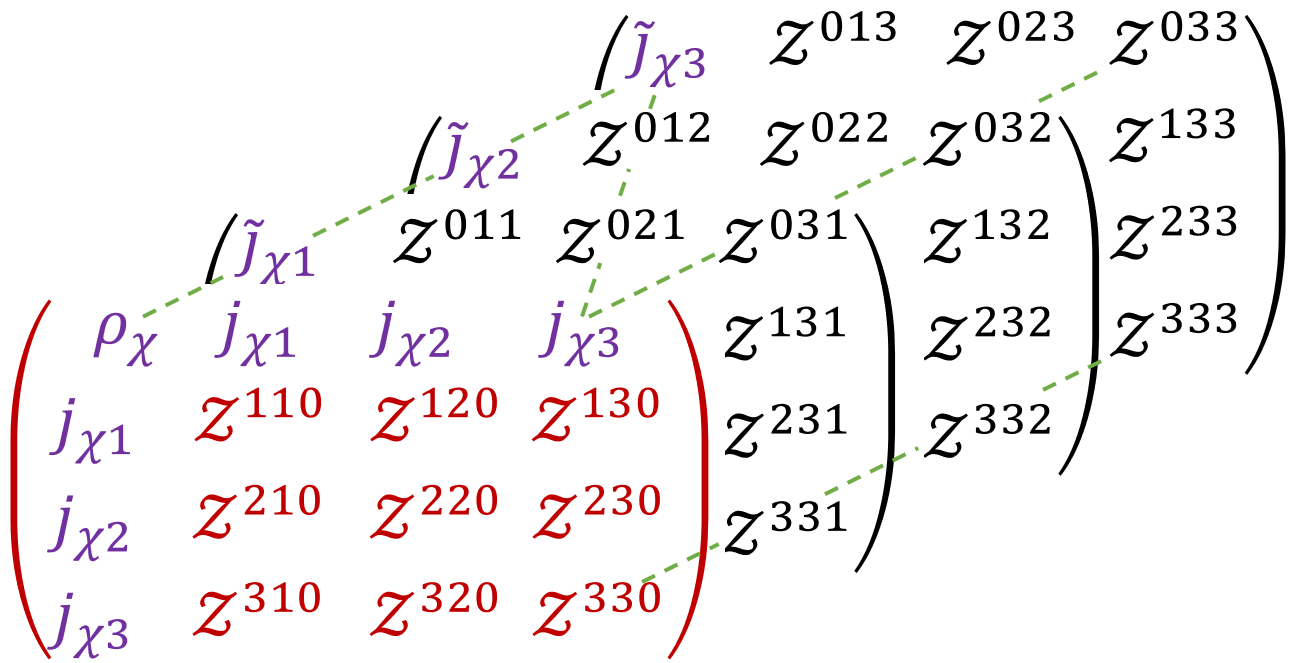}
    \caption{Explicit view of the $Z^{\mu\nu\sigma}$ tensor. The red color denotes the matrix $Z^{ij0}$. The purple components are $Z^{000}=2\rho_\chi$ and $Z^{0i0}=2j_{\chi,i}$ in the normalization adopted here.}
    \label{Explicit view of Z-tensor Fig 2}
\end{figure}

In this section, we review the covariant definition and conservation properties of the Lipkin tensor and isolate the six components $Z^{ab0}$ that will be used throughout the paper. We then express these Lipkin densities and their associated fluxes in a common matrix form. All conservation statements in this section refer to source-free electromagnetic fields in vacuum.

\subsection{Convention}

Greek tensor indices run from 0 to 3 and Latin tensor indices from 1 to 3. We use the Einstein summation convention and the Minkowski metric of signature $g_{\mu\nu}=\mathrm{diag}(+---)$. A spacetime point in standard Minkowski coordinates is denoted by $x^\mu=(t,x^i)$. The totally antisymmetric tensors in four and three dimensions are $\epsilon^{\mu\nu\rho\sigma}$ and $\epsilon^{ijk}$, respectively, with $\epsilon^{0123}=\epsilon_{123}=+1$. We set $c=1$ in Maxwell's equations. Throughout this paper, the upper and lower spatial indices are not distinguished unless explicitly stated otherwise.

Following Lipkin's original component convention~\cite{Lipkin1964}, we represent the electromagnetic field by the antisymmetric tensor
\begin{align}\label{EM tensor def}
F_{\mu\nu}=
\begin{pmatrix}
0 & E_1 & E_2 & E_3\\
-E_1 & 0 & H_3 & -H_2\\
-E_2 & -H_3 & 0 & H_1\\
-E_3 & H_2 & -H_1 & 0
\end{pmatrix}\edit{,}
\end{align}
so that $F_{0i}=E_i$ and $F_{ij}=\epsilon_{ijk}H_k$ in the three-dimensional component notation used throughout this work. With the dual electromagnetic tensor $\tilde{F}^{\mu\nu} = \frac{1}{2}\epsilon^{\mu\nu\alpha\beta} F_{\alpha\beta}$, the source-free Maxwell equations can be expressed compactly in terms of $F$ and $\tilde F$ as

\begin{align}\label{Maxwell eq}
    \begin{split}
        \partial_\mu F^{\mu\nu} = 0,\\
    \partial_\mu \tilde{F}^{\mu\nu} = 0.
    \end{split}
\end{align}
In a covariant form, the zilch is a rank-3 pseudotensor constructed bilinearly from the electromagnetic field and its derivatives. In the following we adopt the original definition proposed by Lipkin. In this convention the zilch tensor has a form~\cite{Lipkin1964}

\begin{align}
\label{Lipkin Tensor def classic}
    Z^{\mu\nu\sigma} = \Bigg[ &\frac{1}{4} g^{\mu\xi}g^{\sigma\delta}\epsilon^{\nu\alpha\gamma\beta} + \frac{1}{4} g^{\nu\xi}g^{\sigma\delta}\epsilon^{\mu\alpha\gamma\beta} + \nonumber\\
    &\frac{1}{4} g^{\mu\xi}g^{\nu\beta}\epsilon^{\sigma\alpha\gamma\delta} + \frac{1}{4} g^{\nu\xi}g^{\mu\beta}\epsilon^{\sigma\alpha\gamma\delta} - \nonumber\\
    &\frac{1}{2} g^{\delta\beta}g^{\mu\gamma}\epsilon^{\nu\sigma\xi\alpha} - \frac{1}{2} g^{\delta\beta}g^{\nu\gamma}\epsilon^{\mu\sigma\xi\alpha} - \\
    &\frac{1}{2} g^{\delta\beta}g^{\mu\alpha}\epsilon^{\nu\sigma\xi\gamma} - \frac{1}{2} g^{\delta\beta}g^{\nu\alpha}\epsilon^{\mu\sigma\xi\gamma}\Bigg] F_{\gamma\delta} \partial_\xi F_{\alpha\beta}\nonumber
\end{align}  
Before algebraic identities are imposed, a rank-three tensor in four dimensions has 64 index components. Here, $\mu,\nu,\sigma=0,1,2,3$ denote one temporal component and three spatial components, $g^{\mu\nu}$ is the metric tensor, $\epsilon^{\mu\nu\sigma\delta}$ is the antisymmetric Levi-Civita tensor, and $F_{\mu\nu}$ is the electromagnetic tensor. With this convention, the $000$ component of the zilch tensor is twice the optical chirality density defined in Eq.~\eqref{OCD def}, $Z^{000}=2\rho_\chi$. A noteworthy property, following directly from the definition~\eqref{Lipkin Tensor def classic}, is the symmetry in the first two indices
\begin{gather}\label{Lipkin Tensor sym classic}
    Z^{\mu\nu\sigma} = Z^{\nu\mu\sigma}.
\end{gather}
Even more importantly, the central result of Lipkin's paper is that this tensor satisfies the divergence equation
\begin{gather}\label{Lipkin Tensor div}
    \partial_\sigma Z^{\mu\nu\sigma}=0.
\end{gather}
For source-free electromagnetic fields, the tensor is also traceless in its first two indices,
\begin{gather}\label{Lipkin Tensor trace}
    g_{\mu\nu}Z^{\mu\nu\sigma}=0.
\end{gather}
Note that the symmetry property~\eqref{Lipkin Tensor sym classic} is off-shell identities, whereas the conservation law~\eqref{Lipkin Tensor div} and the trace identity~\eqref{Lipkin Tensor trace} are on-shell relation that requires the source-free Maxwell equations. The symmetric pair $(\mu,\nu)$ labels ten conserved currents, but the trace identity leaves only nine algebraically independent quantities.

Lipkin identifies $Z^{\mu\nu0}$ as the density of the conserved quantity labeled by the symmetric pair $(\mu,\nu)$ and $Z^{\mu\nu c}$, $c=1,2,3$ as the components of its associated flux.

Equation~\eqref{Lipkin Tensor div} denotes the conservation law in differential form. For the full set of conserved currents, we define
\begin{gather}\label{int density}
    \mathcal Z^{\mu\nu} = \int_V Z^{\mu}\nu0\,d^3x =\mathcal Z^{\nu\mu}
\end{gather}
as the zilch charge contained in the fixed volume $V$ at a given time. Its instantaneous outward flux through the boundary $S=\partial V$ is
\begin{gather}\label{int flux}
    \mathcal F^{\mu\nu} = \int_S Z^{\mu\nu c}\,dS_c = \mathcal F^{\nu\mu}.
\end{gather}
The integral conservation law is therefore
\begin{gather}\label{Lipkin tensor int}
    \frac{d \mathcal Z^{\mu\nu}}{dt} = -\mathcal F^{\mu\nu}.
\end{gather}

\subsection{Density components}

In this subsection, we examine the six components $Z^{ab0}=Z^{ba0}$ of the Lipkin tensor, which we refer to as the Lipkin densities. Starting from Lipkin's explicit expressions, we recast these components as bilinear contractions of the electromagnetic fields with their curls, mediated by six symmetric matrices. This representation reveals the common algebraic structure underlying the Lipkin densities and makes explicit how different pairs of spatial indices select different combinations of field components. It will later provide the link between the Lipkin densities and anisotropic chiral light--matter interaction.

We start with the explicit expressions for density components provided by Lipkin~\cite{Lipkin1964}
\begin{gather}\label{Spatial density componetns}
\boxed{
\begin{split}
    Z^{000} =& \left(\bm E\cdot rot\bm E\right) + \left(\bm H\cdot rot\bm H\right),\\
    Z^{0b0} =& \Bigg[\bm E\times\frac{\partial \bm E}{\partial t} + \bm H\times \frac{\partial \bm H}{\partial t}\Bigg]_{(b)},\\
    Z^{ab0} =& \delta_{ab}\Bigg[\bm E\cdot rot\bm E + \bm H\cdot rot\bm H \Bigg]- \\
    -& E_a (rot \bm E)_b - H_a (rot \bm H)_b - \\
    -& E_b (rot \bm E)_a - H_b (rot \bm H)_a.
\end{split}
}
\end{gather}
The second expression coincides directly with twice the optical-chirality flux defined in Eq.~\eqref{OCF def}, $Z^{0b0}=2j_{\chi,b}$. For source-free electromagnetic fields, Maxwell's equations further establish the equivalence between the first expression and twice the optical chirality density defined in Eq.~\eqref{OCD def}, $Z^{000}=2\rho_\chi$.

The third expression forms a symmetric $3\times3$ matrix of Lipkin densities whose physical interpretation is the central subject of this work
\begin{gather}\label{Spatial density componetns third expression}
\begin{split}
    Z^{ab0} = &\delta_{ab}Z^{000} -  E_a(\bm{\nabla}\times\bm E)_b - E_b(\bm{\nabla}\times\bm E)_a - \\
    -& H_a(\bm{\nabla}\times\bm H)_b - H_b(\bm{\nabla}\times\bm H)_a.
\end{split}
\end{gather}
The first term in Eq.~\eqref{Spatial density componetns third expression} contains the scalar field--curl contraction, whereas the remaining terms resolve bilinear couplings between particular components of the fields and their curls. This structure can be written compactly as
\begin{align}\label{Spatial density componetns third expression 2}
Z^{ab0}
=
\left(\mathcal G^{ab}\right)_{ij}
\left[
E_i(\bm\nabla\times\bm E)_j
+
H_i(\bm\nabla\times\bm H)_j
\right],
\end{align}
where the six symmetric matrices
$\mathcal G^{ab}=\mathcal G^{ba}$ are defined by
\begin{align}\label{Gram matrices compact}
    \left(\mathcal G^{ab}\right)_{ij}
    =
    \delta_{ab}\delta_{ij}
    -
    \delta_{ai}\delta_{bj}
    -
    \delta_{aj}\delta_{bi}.
\end{align}
Here $a,b$ label the Lipkin density, whereas $i,j$ are the contracted indices of the field--curl bilinear. Substitution of Eq.~\eqref{Gram matrices compact} into Eq.~\eqref{Spatial density componetns third expression 2} directly reproduces Eq.~\eqref{Spatial density componetns third expression}.

For $a=b$, Eq.~\eqref{Gram matrices compact} gives a diagonal matrix whose $a$th diagonal element is $-1$, while the other two diagonal elements are $+1$
\begin{align}\label{Gram matrices den 1}
\mathcal{G}^{aa}
=
\begin{pmatrix}
1 - 2\delta_{a1} & 0 & 0 \\
0 & 1 - 2\delta_{a2} & 0 \\
0 & 0 & 1 - 2\delta_{a3}
\end{pmatrix}.
\end{align}
For $a\neq b$, it gives a symmetric off-diagonal matrix with nonzero entries only at the positions $(a,b)$ and $(b,a)$
\begin{align}\label{Gram matrices den 2}
&\mathcal{G}^{ab}
=\nonumber\\
&\begin{pmatrix}
0 &
-\delta_{a1}\delta_{b2}-\delta_{a2}\delta_{b1} &
-\delta_{a1}\delta_{b3}-\delta_{a3}\delta_{b1}
\\
-\delta_{a2}\delta_{b1}-\delta_{a1}\delta_{b2} &
0 &
-\delta_{a2}\delta_{b3}-\delta_{a3}\delta_{b2}
\\
-\delta_{a3}\delta_{b1}-\delta_{a1}\delta_{b3} &
-\delta_{a3}\delta_{b2}-\delta_{a2}\delta_{b3} &
0
\end{pmatrix},\nonumber\\
&
a\neq b.
\end{align}
The six matrices $\mathcal G^{ab}$ are linearly independent and span the six-dimensional vector space of symmetric $3\times3$ matrices. They are not mutually orthogonal with respect to the Frobenius inner product: the three off-diagonal matrices are mutually orthogonal, whereas the three diagonal matrices form a nonorthogonal basis for the diagonal symmetric subspace. This distinction becomes important when the material optical-activity tensor is expanded in the same basis (see App~\ref{App2} for details). The algebraic structure of these matrices is examined in greater detail in Sec.~\ref{sec:Operators of bilinear forms}.

For clarity, we provide the explicit form of Eq.~\eqref{Spatial density componetns third expression 2}
\begin{widetext}
    \begin{gather}\label{Zab0 explicit}
    \begin{split}
        Z^{ab0} = 
        &
        \begin{cases}
            a = b: \left(E_1 \quad E_2 \quad E_3\right)
    \begin{pmatrix}
        1 - 2\delta_{a1} & 0 & 0 \\
        0 & 1 - 2\delta_{a2} & 0 \\
        0 & 0 & 1 - 2\delta_{a3}
    \end{pmatrix}
    \begin{pmatrix}
        (rotE)_1 \\
        (rotE)_2 \\
        (rotE)_3
    \end{pmatrix}\\
            a\neq b: \left(E_1 \quad E_2 \quad E_3\right)
    \begin{pmatrix}
        0 & -\delta_{a1}\delta_{b2} - \delta_{a2}\delta_{b1}  & -\delta_{a1}\delta_{b3} - \delta_{a3}\delta_{b1} \\
        -\delta_{a2}\delta_{b1} - \delta_{a1}\delta_{b2} & 0 & -\delta_{a2}\delta_{b3} - \delta_{a3}\delta_{b2} \\
        -\delta_{a3}\delta_{b1} - \delta_{a1}\delta_{b3} & -\delta_{a3}\delta_{b2} - \delta_{a2}\delta_{b3} & 0
    \end{pmatrix}
    \begin{pmatrix}
        (rotE)_1 \\
        (rotE)_2 \\
        (rotE)_3
    \end{pmatrix}
        \end{cases} +\\
        +
        &
    \begin{cases}
            a = b: \left(H_1 \quad H_2 \quad H_3\right)
    \begin{pmatrix}
        1 - 2\delta_{a1} & 0 & 0 \\
        0 & 1 - 2\delta_{a2} & 0 \\
        0 & 0 & 1 - 2\delta_{a3}
    \end{pmatrix}
    \begin{pmatrix}
        (rotH)_1 \\
        (rotH)_2 \\
        (rotH)_3
    \end{pmatrix}\\
            a\neq b: \left(H_1 \quad H_2 \quad H_3\right)
    \begin{pmatrix}
        0 & -\delta_{a1}\delta_{b2} - \delta_{a2}\delta_{b1}  & -\delta_{a1}\delta_{b3} - \delta_{a3}\delta_{b1} \\
        -\delta_{a2}\delta_{b1} - \delta_{a1}\delta_{b2} & 0 & -\delta_{a2}\delta_{b3} - \delta_{a3}\delta_{b2} \\
        -\delta_{a3}\delta_{b1} - \delta_{a1}\delta_{b3} & -\delta_{a3}\delta_{b2} - \delta_{a2}\delta_{b3} & 0
    \end{pmatrix}
    \begin{pmatrix}
        (rotH)_1 \\
        (rotH)_2 \\
        (rotH)_3
    \end{pmatrix}
        \end{cases}
    \end{split}
    \end{gather}
\end{widetext}

With Lipkin's original definition, the Euclidean trace of the $3\times3$ Lipkin-density block is not constrained to vanish identically. Instead, it satisfies the identity
\begin{gather}\label{trace property}
\sum_{a=1}^{3}Z^{aa0}=Z^{000}.
\end{gather}
For source-free electromagnetic fields, for which $Z^{000}=2\rho_\chi$ in the normalization of Eq.~\eqref{OCD def}, this relation becomes
\begin{gather}
Z^{110}+Z^{220}+Z^{330}=2\rho_\chi.
\end{gather}
This Euclidean trace relation is precisely the $\sigma=0$ component of the four-dimensional Lorentzian trace identity~\eqref{Lipkin Tensor trace},
\begin{align}
    g_{\mu\nu}Z^{\mu\nu0}
    =
    Z^{000}
    -
    \sum_{a=1}^{3}Z^{aa0}
    =
    0.
\end{align}
Thus, the nonvanishing trace allowed for the Lipkin-density block is fully consistent with the identically vanishing Lorentzian trace of the complete Lipkin tensor. The equivalent tensor introduced by Kibble obeys the same trace identity~\cite{kibble1965conservation}.

The expressions above are instantaneous and apply to arbitrary electromagnetic fields that satisfy Maxwell's equations. For comparison with the Tang--Cohen excitation-rate formalism in Sec.~\ref{sec:Zilch tensor and optical activity tensor}, we shall also use their time-averaged monochromatic form. We represent the monochromatic fields as
\begin{align}\label{Monochromatic wave}
\begin{split}
\bm E_{\mathrm{mon}}(\bm r,t)
&=
\operatorname{Re}
\left[
\bm E(\bm r)e^{-i\omega t}
\right],
\\
\bm H_{\mathrm{mon}}(\bm r,t)
&=
\operatorname{Re}
\left[
\bm H(\bm r)e^{-i\omega t}
\right],
\end{split}
\end{align}
where $\bm E(\bm r)$ and $\bm H(\bm r)$ are the corresponding complex field amplitudes. Provided that for two monochromatic quantities with complex amplitudes $A$ and $B$, the average over one optical period is
\begin{align}
    \overline{A_{\mathrm{mon}}B_{\mathrm{mon}}}
    =
    \frac{1}{2}\operatorname{Re}\left(AB^*\right),
\end{align}
It is convenient to display the the time-average of the zeroth slice $Z^{\mu\nu0}$ of the $Z^{\mu\nu\sigma}$ tensor as
\begin{widetext}
\begin{gather}\label{average densities}\begin{split}
\overline{Z^{\mu\nu0}} =
\left(
\begin{array}{cc}
    \omega Im(\bm E \cdot \bm{H^*}) \quad&\quad
    \frac{\omega}{2} Im(\bm{E^*}\times \bm E + \bm{H^*}\times \bm H)_1 \\[0.5em]
    \frac{\omega}{2} Im(\bm{E^*}\times \bm E + \bm{H^*}\times \bm H)_1 \quad&\quad
    \omega Im(-E_1 \cdot H^*_1 + E_2 \cdot H^*_2 + E_3 \cdot H^*_3) \\[0.5em]
    \frac{\omega}{2} Im(\bm{E^*}\times \bm E + \bm{H^*}\times \bm H)_2 \quad&\quad
    -\omega Im(E_1 \cdot H^*_2 + E_2 \cdot H^*_1 ) \\[0.5em]
    \frac{\omega}{2} Im(\bm{E^*}\times \bm E + \bm{H^*}\times \bm H)_3 \quad&\quad
    -\omega Im(E_1 \cdot H^*_3 + E_3 \cdot H^*_1)
\end{array}
\qquad\dots
\right.
\\[1em]
\left.
\begin{array}{cc}
    \frac{\omega}{2} Im(\bm{E^*}\times \bm E + \bm{H^*}\times \bm H)_2 \quad&\quad
    \frac{\omega}{2} Im(\bm{E^*}\times \bm E + \bm{H^*}\times \bm H)_3 \\[0.5em]
    -\omega Im(E_1 \cdot H^*_2 + E_2 \cdot H^*_1) \quad&\quad
    -\omega Im(E_1 \cdot H^*_3 + E_3 \cdot H^*_1) \\[0.5em]
    \omega Im(E_1 \cdot H^*_1 - E_2 \cdot H^*_2 + E_3 \cdot H^*_3) \quad&\quad
    -\omega Im(E_2 \cdot H^*_3 + E_3 \cdot H^*_2) \\[0.5em]
    -\omega Im(E_2 \cdot H^*_3 + E_3 \cdot H^*_2) \quad&\quad
    \omega Im(E_1 \cdot H^*_1 + E_2 \cdot H^*_2 - E_3 \cdot H^*_3)
\end{array}
\right)
\end{split}
\end{gather}
\end{widetext}

This $4\times4$ symmetric density slice contains ten unique components, nine of which are algebraically independent because of Eq.~\eqref{Lipkin Tensor trace}. The temporal row and column contain $Z^{000}=2\rho_\chi$ and $Z^{0i0}=2j_{\chi,i}$, while the $3\times3$ block consists of the Lipkin densities. We will use Eqs.~\eqref{Spatial density componetns third expression 2},~\eqref{Zab0 explicit}, and~\eqref{average densities} in Sec.~\ref{sec:Zilch tensor and optical activity tensor}, where the anisotropic molecular response is considered.

\subsection{Flux components}

The continuity equation~\eqref{Lipkin Tensor div} associates each density component $Z^{\mu\nu0}$ with a corresponding flux $Z^{\mu\nu c}$, $c=1,2,3$. In this subsection we examine the explicit expressions originally derived by Lipkin~\cite{Lipkin1964} and show that the fluxes $Z^{abc}$ associated with the six Lipkin densities $Z^{ab0}$ are governed by the same symmetric matrices $\mathcal G^{ab}$ as the densities themselves.

We now turn to the explicit expressions for the flux components originally derived by Lipkin~\cite{Lipkin1964}
\begin{gather}\label{flux components}
\boxed{
\begin{split}
    Z^{00c} =& Z^{0c0} - \left(rot(\bm E\times\bm H)\right)_{c},\\
    Z^{0bc} =& Z^{bc0} + \epsilon^{cdp} \Bigg[ \frac{1}{2} \left(\bm E^2 + \bm H^2\right)\delta_{bd} - \\
    -& E_b E_d + H_b H_d\Bigg]_{(,p)},\\
    Z^{abc} =& \delta_{ab} Z^{00c} + H_a E_{b,c} - E_a H_{b,c} + \\
    +& H_{b} E_{a,c} - E_b H_{a,c}.
\end{split}
}
\end{gather}
Here, the following notation is used
\begin{gather*}
    F_{i,k} = \frac{\partial F_i}{\partial x_k}, \quad \bm F = \bm E,\bm H.
\end{gather*}
In the analysis of Eq.~\eqref{flux components}, we will not examine the second expression in detail and will instead restrict ourselves to the internal components of the tensor, namely the first and third expressions. While the matrix $Z^{0bc}$ is of great interest for problems involving anisotropy of the chirality flux, in this paper we focus on the remaining components.

We first consider the flux components $Z^{00c}$, which are related to $Z^{0c0}$ by
\begin{gather}\label{flux component 1}
Z^{00c}
=
Z^{0c0}
-
\left[
\bm\nabla\times(\bm E\times\bm H)
\right]_c.
\end{gather}
The components $Z^{00c}$ form the flux associated with the conserved density $Z^{000}$ in the Lipkin current $Z^{00\sigma}$. Since $Z^{0c0}=2j_{\chi c}$, the flux associated directly with this current can be defined as
\begin{align}
    \widetilde{\bm j}_{\chi}
    &=
    \frac{1}{2}
\begin{pmatrix}
    Z^{001}\\
    Z^{002}\\
    Z^{003}
\end{pmatrix}
    =
    \bm j_\chi
    -
    \frac{1}{2}
    \bm\nabla\times(\bm E\times\bm H).
\label{Chiral flux tilde}
\end{align}
Thus, $\widetilde{\bm j}_\chi$ and the conventionally defined optical-chirality flux $\bm j_\chi$ differ only by a curl term. Together with the source-free Maxwell equations, it gives
\begin{align}
\widetilde j_{\chi c}
    =
    \frac{1}{2}Z^{00c}
    =
    \frac{1}{2}
    \left(
    E_i\partial_c H_i
    -
    H_i\partial_c E_i
    \right).
\label{Chiral flux tilde components}
\end{align}
Equivalently,
\begin{align}
\widetilde{\bm j}_{\chi}
=
\frac{1}{2}
\begin{pmatrix}
\bm E\cdot\partial_1\bm H-\bm H\cdot\partial_1\bm E\\
\bm E\cdot\partial_2\bm H-\bm H\cdot\partial_2\bm E\\
\bm E\cdot\partial_3\bm H-\bm H\cdot\partial_3\bm E
\end{pmatrix}.
\end{align}
Since the divergence of a curl vanishes identically, the two fluxes have the same divergence. The conservation law for the current $Z^{00\sigma}$ therefore reproduces the optical-chirality continuity equation
\begin{align}
\partial_\sigma Z^{00\sigma}
&=
\partial_0Z^{000}
+
\partial_cZ^{00c} =
\nonumber\\
&=
2
\left(
\frac{\partial\rho_\chi}{\partial t}
+
\bm\nabla\cdot\widetilde{\bm j}_\chi
\right)
=
2
\left(
\frac{\partial\rho_\chi}{\partial t}
+
\bm\nabla\cdot\bm j_\chi
\right)
=
0.
\end{align}
The position of these components within the Lipkin tensor is shown in Fig.~\ref{Explicit view of Z-tensor Fig 2}.

We now turn to the flux components $Z^{abc}$. The third expression in Eq.~\eqref{flux components} relates them to the optical-chirality flux column $\widetilde j_{\chi c}$
\begin{gather}\label{flux components 3}
\begin{split}
Z^{abc}
={}&
2\delta_{ab}\widetilde j_{\chi c}
-
E_a\partial_cH_b
-
E_b\partial_cH_a
\\
&{}
+
H_a\partial_cE_b
+
H_b\partial_cE_a.
\end{split}
\end{gather}
The same matrices $\mathcal G^{ab}$ that appear in the Lipkin densities also determine the structure of these fluxes. Using Eq.~\eqref{Gram matrices compact}, we obtain
\begin{align}\label{flux components 3 gram}
Z^{abc}
=
\left(\mathcal G^{ab}\right)_{ij}
\left(
E_i\partial_cH_j
-
H_i\partial_cE_j
\right).
\end{align}
Thus, the density components $Z^{ab0}$ and their associated fluxes $Z^{abc}$ are generated by the same six symmetric matrices $\mathcal G^{ab}$. This common algebraic structure motivates the detailed analysis of these matrices in Sec.~\ref{sec:Operators of bilinear forms}.

For clarity, we provide the explicit form of Eq.~\eqref{flux components 3 gram} and the time-averaged value of the $\sigma$-th slice of the zilch tensor $Z^{\mu\nu c}$
\begin{widetext}
\begin{gather}\label{Explicit view Zabc}
    \begin{split}
        Z^{abc} = 
        &
        \begin{cases}
            a = b: \left(E_1 \quad E_2 \quad E_3\right)
    \begin{pmatrix}
        1 - 2\delta_{a1} & 0 & 0 \\
        0 & 1 - 2\delta_{a2} & 0 \\
        0 & 0 & 1 - 2\delta_{a3}
    \end{pmatrix}
    \partial_c
    \begin{pmatrix}
        H_1 \\
        H_2 \\
        H_3
    \end{pmatrix}\\
            a\neq b: \left(E_1 \quad E_2 \quad E_3\right)
    \begin{pmatrix}
        0 & -\delta_{a1}\delta_{b2} - \delta_{a2}\delta_{b1}  & -\delta_{a1}\delta_{b3} - \delta_{a3}\delta_{b1} \\
        -\delta_{a2}\delta_{b1} - \delta_{a1}\delta_{b2} & 0 & -\delta_{a2}\delta_{b3} - \delta_{a3}\delta_{b2} \\
        -\delta_{a3}\delta_{b1} - \delta_{a1}\delta_{b3} & -\delta_{a3}\delta_{b2} - \delta_{a2}\delta_{b3} & 0
    \end{pmatrix}
    \partial_c
    \begin{pmatrix}
        H_1 \\
        H_2 \\
        H_3
    \end{pmatrix}
        \end{cases} -\\
        -
        &
    \begin{cases}
            a = b: \left(H_1 \quad H_2 \quad H_3\right)
    \begin{pmatrix}
        1 - 2\delta_{a1} & 0 & 0 \\
        0 & 1 - 2\delta_{a2} & 0 \\
        0 & 0 & 1 - 2\delta_{a3}
    \end{pmatrix}
    \partial_c
    \begin{pmatrix}
        E_1 \\
        E_2 \\
        E_3
    \end{pmatrix}\\
            a\neq b: \left(H_1 \quad H_2 \quad H_3\right)
    \begin{pmatrix}
        0 & -\delta_{a1}\delta_{b2} - \delta_{a2}\delta_{b1}  & -\delta_{a1}\delta_{b3} - \delta_{a3}\delta_{b1} \\
        -\delta_{a2}\delta_{b1} - \delta_{a1}\delta_{b2} & 0 & -\delta_{a2}\delta_{b3} - \delta_{a3}\delta_{b2} \\
        -\delta_{a3}\delta_{b1} - \delta_{a1}\delta_{b3} & -\delta_{a3}\delta_{b2} - \delta_{a2}\delta_{b3} & 0
    \end{pmatrix}
    \partial_c
    \begin{pmatrix}
        E_1 \\
        E_2 \\
        E_3
    \end{pmatrix}
        \end{cases}
    \end{split}
\end{gather}
\begin{gather}\label{average fluxes}\begin{split}
\overline{Z^{\mu\nu c}} =
\left(
\begin{array}{cc}
    \frac{1}{2} Re(\bm E \cdot \bm{H^*}_{,\sigma} - \bm H \cdot \bm{E^*}_{,\sigma}) \quad&\quad
    \overline{Z^{01\sigma}} \\[0.5em]
    \overline{Z^{10\sigma}} \quad&\quad
    \frac{1}{2} Re(\bm E \cdot \bm{H^*}_{,\sigma} - \bm H \cdot \bm{E^*}_{,\sigma} + 2(H_1E^*_{1,\sigma}-E_1H^*_{1,\sigma})) \\[0.5em]
    \overline{Z^{20\sigma}} \quad&\quad
    \frac{1}{2} Re(H_2E^*_{1,\sigma} + H_1E^*_{2,\sigma} - E_2H^*_{1,\sigma} - E_1H^*_{2,\sigma})\\[0.5em]
    \overline{Z^{30\sigma}} \quad&\quad
    \frac{1}{2} Re(H_3E^*_{1,\sigma} + H_1E^*_{3,\sigma} - E_3H^*_{1,\sigma} - E_1H^*_{3,\sigma})
\end{array}
\qquad\dots
\right.
\\[1em]
\left.
\begin{array}{cc}
    \overline{Z^{02\sigma}} \quad&\quad
    \overline{Z^{03\sigma}} \\[0.5em]
    \frac{1}{2} Re(H_1E^*_{2,\sigma} + H_2E^*_{1,\sigma} - E_1H^*_{2,\sigma} - E_2H^*_{1,\sigma}) \quad&\quad
    \frac{1}{2} Re(H_3E^*_{1,\sigma} + H_1E^*_{3,\sigma} - E_3H^*_{1,\sigma} - E_1H^*_{3,\sigma}) \\[0.5em]
    \frac{1}{2} Re(\bm E \cdot \bm{H^*}_{,\sigma} - \bm H \cdot \bm{E^*}_{,\sigma} + 2(H_2E^*_{2,\sigma}-E_2H^*_{2,\sigma})) \quad&\quad
    \frac{1}{2} Re(H_2E^*_{3,\sigma} + H_3E^*_{2,\sigma} - E_2H^*_{3,\sigma} - E_3H^*_{2,\sigma}) \\[0.5em]
    \frac{1}{2} Re(H_3E^*_{2,\sigma} + H_2E^*_{3,\sigma} - E_3H^*_{2,\sigma} - E_2H^*_{3,\sigma}) \quad&\quad
    \frac{1}{2} Re(\bm E \cdot \bm{H^*}_{,\sigma} - \bm H \cdot \bm{E^*}_{,\sigma} + 2(H_3E^*_{3,\sigma}-E_3H^*_{3,\sigma}))
\end{array}
\right)
\end{split}
\end{gather}
\end{widetext}

Summarizing this section, the six Lipkin densities $Z^{ab0}$ and their associated flux components $Z^{abc}$ allow a common matrix representation, Eqs.~\eqref{Spatial density componetns third expression 2} and~\eqref{flux components 3 gram}. These matrices are analyzed in Sec.~\ref{sec:Operators of bilinear forms} and subsequently provide the algebraic basis for relating the Lipkin densities to anisotropic optical activity in Sec.~\ref{sec:Zilch tensor and optical activity tensor}.

\section{\label{sec:Operators of bilinear forms}Algebraic structure of the Lipkin densities}

In this section we analyze the matrices $\mathcal{G}$ introduced in
Eqs.~\eqref{Gram matrices den 1}–\eqref{Gram matrices den 2}. 
These matrices appear in the representation of the Lipkin densities $Z^{ab0}$ and play the role of algebraic operators acting on the bilinear combinations of the electromagnetic field.

For convenience, we separate the six matrices into two groups. The three diagonal matrices are denoted by $A^{(m)}$ and the three off-diagonal matrices by $B^{(m)}$, with $m=1,2,3$. Their explicit form is

\begin{equation}\label{Matrices A and B}
\begin{alignedat}{2}
A^{(1)} &= \mathcal{G}^{11} =
\begin{pmatrix}
-1 & 0 & 0 \\
0 & 1 & 0 \\
0 & 0 & 1
\end{pmatrix}, \quad&
B^{(1)} &= \mathcal{G}^{12} =
\begin{pmatrix}
0 & -1 & 0 \\
-1 & 0 & 0 \\
0 & 0 & 0
\end{pmatrix}, \\
A^{(2)} &= \mathcal{G}^{22} =
\begin{pmatrix}
1 & 0 & 0 \\
0 & -1 & 0 \\
0 & 0 & 1
\end{pmatrix}, \quad&
B^{(2)} &= \mathcal{G}^{13} =
\begin{pmatrix}
0 & 0 & -1 \\
0 & 0 & 0 \\
-1 & 0 & 0
\end{pmatrix}, \\
A^{(3)} &= \mathcal{G}^{33} =
\begin{pmatrix}
1 & 0 & 0 \\
0 & 1 & 0 \\
0 & 0 & -1
\end{pmatrix}, \quad&
B^{(3)} &= \mathcal{G}^{23} =
\begin{pmatrix}
0 & 0 & 0 \\
0 & 0 & -1 \\
0 & -1 & 0
\end{pmatrix}.
\end{alignedat}
\end{equation}

All matrices in Eq.~\eqref{Matrices A and B} are symmetric and linearly independent, and together they form a convenient basis of the six-dimensional vector space of symmetric $3\times3$ matrices. They should not be confused with an orthonormal traceless basis: the matrices $A^{(m)}$ have nonzero trace and are not mutually orthogonal, whereas the matrices $B^{(m)}$ are off-diagonal and mutually orthogonal.

The matrices $A^{(m)}$ generate the three diagonal Lipkin densities from linear combinations of the diagonal components of $K_{ij}$, whereas the matrices $B^{(m)}$ extract its symmetric off-diagonal combinations. In the present problem, the relevant $3\times3$ bilinear object is the field--curl pseudotensor
\begin{align}
K_{ij}
=
E_i(\nabla\times \mathbf E)_j
+
H_i(\nabla\times \mathbf H)_j .
\label{eq:field_curl_tensor}
\end{align}

The matrices in Eq.~\eqref{Matrices A and B} act as algebraic operators that contract this tensor with particular symmetric combinations of Cartesian indices. They are not projectors in the strict algebraic sense; rather, they provide a Lipkin-adapted basis of linear contractions. In this notation, the Lipkin densities can be written compactly as
\begin{align}
Z^{ab0}
=
(\mathcal G^{ab})_{ij}K_{ij}\edit{.}
\label{eq:Bilinear form}
\end{align}
Explicitly, Eq.~\eqref{eq:Bilinear form} is equivalent to
\begin{align}
Z^{ab0}
=
\delta_{ab}K_{kk}
-
K_{ab}
-
K_{ba}.
\label{eq:Zab0_from_K}
\end{align}

Equation~\eqref{eq:Zab0_from_K} makes it clear why only the symmetric part of $K_{ij}$ contributes to the Lipkin-density block. Indeed, decomposing $K_{ij}$ into symmetric and antisymmetric parts 
\begin{align}
\begin{split}
&K_{ij}=K_{(ij)}+K_{[ij]}, \\
&K_{(ij)}=\frac12(K_{ij}+K_{ji}),
\quad
K_{[ij]}=\frac12(K_{ij}-K_{ji}),
\label{eq:SO3_1}
\end{split}
\end{align}
one immediately obtains
\begin{align}
Z^{ab0}
=
\delta^{ab}K_{kk}
-
2K_{(ab)}.
\label{eq:Zab0_from_S}
\end{align}
The antisymmetric part $K_{[ij]}$ cancels identically. Thus, although a general second-rank pseudotensor $K_{ij}$ contains symmetric and antisymmetric parts, the Lipkin densities $Z^{ab0}=Z^{ba0}$ are built entirely from the symmetric part of the field--curl pseudotensor $K_{ij}$.

This observation is naturally expressed in terms of the irreducible decomposition of a second-rank tensor under spatial rotations,
\begin{align}
\mathbf 3\otimes \mathbf 3
=
\mathbf 1\oplus \mathbf 3\oplus \mathbf 5.
\label{eq:SO3_decomp}
\end{align}
Here $\mathbf 1$ corresponds to the one-dimensional trace part, $\mathbf 3$ to the antisymmetric part, and $\mathbf 5$ to the symmetric-traceless part. Since the Lipkin-density block is symmetric in the indices $a,b$, the antisymmetric $\mathbf 3$ sector is absent in Eq.~\eqref{eq:Zab0_from_S}. The six components $Z^{ab0}$ therefore decompose into an isotropic trace contribution and a five-dimensional symmetric-traceless contribution. The isotropic contribution is fixed by the trace of the block, which is equal to $Z^{000}=2\rho_\chi$ for source-free fields, whereas the remaining five independent combinations constitute the anisotropic tensorial part.

It remains useful to indicate what happens to the antisymmetric sector that is absent from Eq.~\eqref{eq:Zab0_from_S}. For the field--curl tensor $K_{ij}$, the antisymmetric part is
\begin{align}
\begin{split}
K_{[ij]}
=
\frac12
\Bigg[
E_i(\nabla\times\mathbf E)_j
-
E_j(\nabla\times\mathbf E)_i +\\
+
H_i(\nabla\times\mathbf H)_j
-
H_j(\nabla\times\mathbf H)_i
\Bigg].
\label{eq:antisymmetric_K}
\end{split}
\end{align}
Using the duality between antisymmetric rank-two pseudotensors and polar vectors, $K_{[ij]}=\epsilon_{ijk}v_k$, one obtains
\begin{align}
v_k
=
\frac12
\left[
\mathbf E\times(\nabla\times\mathbf E)
+
\mathbf H\times(\nabla\times\mathbf H)
\right]_k .
\label{eq:antisymmetric_dual_vector}
\end{align}
For source-free fields this vector is related to the divergence of the Maxwell stress tensor
\begin{align}
\sigma_{ij}
=
E_iE_j+H_iH_j
-
\frac12\delta_{ij}(E^2+H^2)
\label{eq:Maxwell_stress}
\end{align}
through
\begin{align}
\partial_j\sigma_{ij}
=
-
\left[
\mathbf E\times(\nabla\times\mathbf E)
+
\mathbf H\times(\nabla\times\mathbf H)
\right]_i
=
-2v_i .
\label{eq:antisymmetric_stress_relation}
\end{align}
Thus, the antisymmetric $\mathbf 3$ sector does not contribute to the symmetric Lipkin-density block and is instead expressible through the divergence of the Maxwell stress tensor.

\section{\label{sec:Zilch tensor and optical activity tensor}Anisotropic optical activity}

As mentioned, Tang and Cohen demonstrated~\cite{Tang2010} that the excitation-rate difference for the opposite enantiomers of a small isotropic chiral molecule in the electric-dipole--magnetic-dipole approximation is governed by the optical chirality density, which in the present normalization equals $Z^{000}/2$. In what follows, we extend this result to the general reciprocal bi-anisotropic molecule. We will show that the Lipkin densities $Z^{ab0}$ emerge naturally as the field quantities coupled to the symmetric part of the optical-activity tensor.

Considering a general monochromatic electromagnetic field~\eqref{Monochromatic wave} and  following the approach of Tang and Cohen~\cite{Tang2010} for the modeling of light-matter interaction, we use an expression for the molecular excitation rate averaged over one optical period of the electromagnetic field
\begin{align}\label{Tang velocity}
    A^\pm = \langle \mathbf{E}\cdot\dot{\mathbf{p}} + \mathbf{H}\cdot\dot{\mathbf{m}}\rangle = \frac{\omega}{2}Im(\mathbf{E}^*\cdot\mathbf{p} + \mathbf{H}^*\cdot\mathbf{m})
\end{align}
together with constitutive relations for the induced electric and magnetic dipole moments. In the isotropic case, these relations take the form \cite{Fedorov, Barron}
\begin{align}\label{Tang dipoles iso}
    \mathbf{p}=\alpha\mathbf{E} - iG\mathbf{H},\quad \mathbf{m}=\chi\mathbf{H} + iG\mathbf{E}.
\end{align}
Here $\alpha$ is electric polarizability, $\chi$ is magnetic susceptibility, and $G$ is isotropic mixed magneto-electric dipole polarizability (optical activity material parameter); all of these parameters have real and imaginary parts:
\begin{align}\label{Real and Imag}
    f = f' + if'',\quad f = \alpha,\chi,G,
\end{align}
Substituting~\eqref{Tang dipoles iso} into~\eqref{Tang velocity} yields the Tang-Cohen expression
\begin{align}\label{Tang velocity iso}
    A^\pm_{iso} = \frac{\omega}{2}\left(\alpha''|\mathbf{E}|^2 + \chi''|\mathbf{H}|^2\right) \pm G''\omega Im(\bm{E}^*\cdot\bm{H}).
\end{align}
The last term is proportional (up to a sign convention) to the optical chirality density~\eqref{OCD def}, averaged over one period. Tang and Cohen therefore concluded that, in the isotropic case, the chiral asymmetry in the molecular excitation rate is proportional to the product of the molecular chirality and the chirality of the electromagnetic field~\cite{Tang2010}.


To extend this analysis to general reciprocal bi-anisotrpic response, we use the following reciprocal constitutive relations ~\cite{Fedorov,Barron,YangCohen2011}
\begin{align}\label{Tang dipoles ani}
\begin{split}
p_a^\pm = \alpha_{ab}E_b \mp iG_{ab}H_b,\qquad m_a^\pm = \chi_{ab}H_b \pm iG_{ba}E_b,
\end{split}
\end{align}
where $G_{ab}$ is a rank-two magneto-electric pseudotensor determined by the geometry and symmetry of the material response~\cite{Fedorov,Barron,LandauVol8}. The reciprocity also implies that $\alpha_{ab}=\alpha_{ba}$ and $\chi_{ab}=\chi_{ba}$. The negative transpose of $G_{ab}$ in Eq.~\eqref{Tang dipoles ani} follows from the reciprocity of the mixed electric--magnetic response. The signs $\pm$ represent two responses related by $G_{ab}\rightarrow-G_{ab}$, with $\alpha_{ab}$ and $\chi_{ab}$ kept fixed. In case of chiral molecules, this corresponds to two opposite material enantiomers.

\begin{table*}[t]
\centering
\renewcommand{\arraystretch}{1.25}
\begin{tabular}{p{0.10\textwidth}|p{0.48\textwidth}|p{0.18\textwidth}|p{0.20\textwidth}}
\hline
Symmetry group & Symmetry operations & Allowed nonzero components of $G''_{(ab)}$&{Class of medium}\\
\hline
$C_1$ 
& $E$ 
& all components {$G''_{(ab)}$} & {general bi-anisotropic}\\
\hline
$C_2$ 
& $E$, $C_2(z)$ 
& {$G''_{xx},G''_{yy},G''_{zz},G''_{xy}$}& {anisotropic chiral}\\
\hline
$C_s$ 
& $E$, $\sigma_h(xy)$ 
& {$G''_{xz},G''_{yz}$}& {pseudochiral}
\\
\hline
$C_{2v}$ 
& $E$, $C_2(z)$, $\sigma_v(xz)$, $\sigma_v(yz)$ 
& {$G''_{xy}$} & {pseudochiral}
\\
\hline
$D_2$ 
& $E$, $C_2(x)$, $C_2(y)$, $C_2(z)$ 
&{$G''_{xx},G''_{yy},G''_{zz}$} & {anisotropic chiral}
\\
\hline
{$C_n$ ($n\geq3$)}
& $E$, $C_n(z)$, $C_n^2(z),\ldots,C_n^{n-1}(z)$ 
& {$G''_{xx}=G''_{yy},G''_{zz}$} & {anisotropic chiral}
\\
\hline
{$D_n$ ($n\geq3$)}
& $E$, $C_n^k(z)$, and $n$ twofold rotations $C_2'$ about axes in the $xy$ plane 
& {$G''_{xx}=G''_{yy},G''_{zz}$} & {anisotropic chiral}
\\
\hline
$S_4$ 
& $E$, $S_4(z)$, $C_2(z)=S_4^2$, $S_4^3(z)$ 
& {$G''_{xx}=-G''_{yy},G''_{xy}$} & {anisotropic chiral}
\\
\hline
$D_{2d}$ 
& $E$, $C_2(z)$, two $C_2'$ rotations, two $S_4$ operations, and two diagonal mirror planes $\sigma_d$ 
& {$G''_{xy}$} & {pseudochiral}
\\
\hline
$T,O$ 
& proper rotations of the tetrahedron or cube/octahedron 
& {$G''_{xx}=G''_{yy}=G''_{zz}$} & {isotropic chiral}
\\
\hline
\end{tabular}
\caption{Relation between selected point groups, their symmetry operations, and the allowed nonzero components of the symmetric optical-activity pseudotensor $G''_{(ab)}$. The principal symmetry axis is chosen along $z$; for $C_s$, the mirror plane is the $xy$ plane. The rows $C_n$ and $D_n$ with $n\geq3$ include noncrystallographic molecular groups such as $C_5$ and $D_5$.}
\label{tab:Gsym}
\end{table*}



Next, following the procedure used in classification of bi-anisotropic media \cite{tretyakov1998magnetoelectric}, we decompose the general reciprocal pseudotensor $G_{ab}$ into diagonal and traceless symmetric and antisymmetric parts
\begin{align}\label{eq:G_Tretyakov_decomposition}
    G_{ab}
    =
    G\delta_{ab}
    +
    G_{(ab)}
    +
    G_{[ab]},
\end{align}
where
\begin{align}
\begin{split}
    &G
    =
    \frac13 G_{aa},
    \\
    &G_{(ab)}
    =
    \frac12
    \left(
        G_{ab}
        +
        G_{ba}
    \right)
    -
    G\delta_{ab},
    \\
    &G_{[ab]}
    =
    \frac12
    \left(
        G_{ab}
        -
        G_{ba}
    \right)
    =
    \epsilon_{abc}\Omega_c.
\end{split}
\end{align}
Here $G = \frac13\mathrm{tr}\{G_{ab}\}$ is the pseudoscalar magneto-electric parameter,  $G_{(ab)}$ denotes symmetric traceless part of $G_{ab}$ pseudotensor, and $G_{[ab]}$ is the antisymmetric traceless part. Note that the latter can always be equivalently represented via the polar vector \editS{ $\bm\Omega = \frac{1}{2}\epsilon_{abc}G_{ab} =  [G_{[23]}, -G_{[13]}, G_{[12]}]$.} 

In this terminology, the first term in \eqref{eq:G_Tretyakov_decomposition} describes isotropic chiral medium (Pasteur medium), the second term corresponds to anisotropic chiral or pseudochiral medium, while the last term represents so-called omega medium. All terms together describe general bi-anisotropic medium.


Substitution of reciprocal constitutive relations~\eqref{Tang dipoles ani} into~\eqref{Tang velocity} gives
\begin{align}\label{eq:A_ani_before_decomposition}
    A^\pm_{ani} = A_0 \pm \omega G_{ab}'' \operatorname{Im} \left( E_a^*H_b \right),
\end{align}
where
\begin{align}
A_0 = \frac{\omega}{2} \left[\alpha''_{ab}\operatorname{Re} \left(E_a^*E_b\right) + \chi''_{ab}\operatorname{Re} \left(H_a^*H_b\right) \right].
\end{align}
Using Eq.~\eqref{eq:G_Tretyakov_decomposition}, we separate the second term in Eq.~\eqref{eq:A_ani_before_decomposition} into three independent contributions
\begin{align}\label{eq:A_mix_three_sectors}
\begin{split}
    A^\pm_{ani}
    =A_0
    {}&
    \pm
    \omega G''
    \delta_{ab}
    \operatorname{Im}
    \left(
        E_a^*H_b
    \right)
    \\
    &\pm
    \omega G_{(ab)}''
    \operatorname{Im}
    \left(
        E_a^*H_b
    \right)
    \\
    &\pm
    \omega G_{[ab]}''
    \operatorname{Im}
    \left(
        E_a^*H_b
    \right) = \\
    & = A_0 + A_{iso}^\pm + A_{pseudo}^\pm + A_{omega}^\pm.
\end{split}
\end{align}
In Eq.~\eqref{eq:A_mix_three_sectors}, the trace part, as well as the symmetric and antisymmetric parts of $G''_{ab}$ are coupled only with the corresponding parts of $\operatorname{Im}(E_a^*H_b)$ tensor. Thus,
\begin{align}
    G_{(ab)}''
    \operatorname{Im}
    \left(
        E_a^*H_b
    \right)
    =
    \frac12
    G_{(ab)}''
    \operatorname{Im}
    \left(
        E_a^*H_b
        -
        H_a^*E_b
    \right),
\end{align}
whereas
\begin{align}
\begin{split}
    G_{[ab]}''
    \operatorname{Im}
    \left(
        E_a^*H_b
    \right)
    &=
    \epsilon_{abc}\Omega_c''
    \operatorname{Im}
    \left(
        E_a^*H_b
    \right)
    \\
    &=
    \bm\Omega''
    \cdot
    \operatorname{Im}
    \left(
        \mathbf E^*
        \times
        \mathbf H
    \right).
\end{split}
\end{align}
Therefore
\begin{align}\label{eq:A_mix_decomposed_field}
\begin{split}
    A^\pm_{ani}
    = A_0
    {}&
    \pm
    \omega G''
    \operatorname{Im}
    \left(
        \mathbf E^*
        \cdot
        \mathbf H
    \right)
    \\
    &\pm
    \frac{\omega}{2}
    G_{(ab)}''
    \operatorname{Im}
    \left(
        E_a^*H_b
        -
        H_a^*E_b
    \right)
    \\
    &\pm
    \omega
    \bm\Omega''
    \cdot
    \operatorname{Im}
    \left(
        \mathbf E^*
        \times
        \mathbf H
    \right).
\end{split}
\end{align}
The first term in \eqref{eq:A_mix_decomposed_field} is the isotropic Tang--Cohen term and can be expressed via the Lipkin tensor as 
\begin{align}
    A^\pm_{iso}
    =
    \pm
    \omega G''
    \operatorname{Im}
    \left(
        \mathbf E^*
        \cdot
        \mathbf H
    \right)
    =
    \mp
    G''
    \overline{Z^{000}} .
\end{align}

The second term represents the symmetric traceless (i.e., pseudochiral anisotropic) contribution: since $G_{(ab)}$ is traceless, it does not couple to the optical chirality density but instead to the traceless anisotropic part of the Lipkin density block (see App.~\ref{App1} for details). The pseudochiral contribution to the excitation rate is therefore
\begin{align}\label{eq:A_pseudochiral_sector}
    A^\pm_{pseudo}
    =
    \pm
    \frac{\omega}{2}
    G_{(ab)}''
    \operatorname{Im}
    \left(
        E_a^*H_b
        -
        H_a^*E_b
    \right)
    =
    \pm
    \frac12
    G_{(ab)}''
    \overline{Z^{(ab)0}}.
\end{align}

The third term in~\eqref{eq:A_mix_decomposed_field} is qualitatively different. It refers to the reciprocal omega-type magneto-electric coupling. Its field combination \edit{$\operatorname{Im}\left(\mathbf E^*\times   \mathbf H\right)$} is associated with the reactive part of \edit{the complex Poynting vector} rather than with the Lipkin density block. Therefore the antisymmetric part of the magneto-electric pseudotensor is not coupled directly with any part of the Lipkin tensor.

Combining all three contributions, we obtain
\begin{align}\label{eq:Central_Tretyakov_Lipkin_result}
\boxed{
    A^\pm_{ani}
    =
    A_0
    \mp
    G''
    \overline{Z^{000}}
    \pm
    \frac12
    G_{(ab)}''
    \overline{Z^{(ab)0}}
    \pm
    \omega
    \bm\Omega''
    \cdot
    \operatorname{Im}
    \left(
        \mathbf E^*
        \times
        \mathbf H
    \right)
}
\end{align}
Equation~\eqref{eq:Central_Tretyakov_Lipkin_result} separates the
reciprocal mixed electric--magnetic response into three different parts. The trace part $G''$ couples to the Lipkin density $\overline{Z^{000}}$, reproducing the Tang--Cohen result. The symmetric traceless part $G_{(ab)}''$ couples to the traceless symmetric anisotropic Lipkin-density block $\overline{Z^{(ab)0}}$. The antisymmetric part, represented by $\bm\Omega''$, gives the reciprocal omega-type contribution and is conjugate to $\operatorname{Im}(\mathbf E^*\times\mathbf H)$ rather than to Lipkin densities.

Finally, Table~\ref{tab:Gsym} summarizes the relation between selected molecular point groups, the symmetry-allowed nonzero components of $G''_{(ab)}$ \cite{LandauVol8} and the class of bi-anisotropic material according to classification in \cite{tretyakov1998magnetoelectric}. The component lists are written in the conventional coordinate frame adapted to the symmetry axes of each point group. Note that for some of the materials in Table~\ref{tab:Gsym}, the net chirality vanishes; they are referred to as pseudochiral.

\section{\label{sec:Numerical example}Numerical example}

\begin{figure*}[t]
    \centering
    \includegraphics[width=0.7\linewidth]{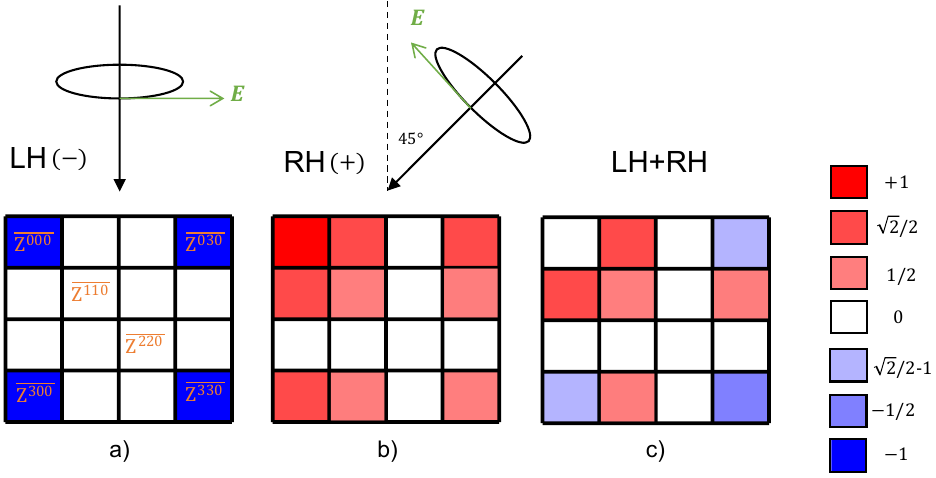}
    \caption{Emergence of purely anisotropic optical chirality under LH--RH superposition: calculated time-averaged density slice $\overline{Z^{\mu\nu0}}$ for (a) a left-handed circularly polarized wave propagating along the $z$ axis; (b) a right-handed circularly polarized wave propagating in the $xz$ plane at an angle $\pi/4$ with respect to the $z$ axis; and (c) the superposition of the two monochromatic waves. Indices $\mu=\nu=0$ are indicated in the upper-left corner of the tables. The index $\mu$ corresponds to rows and $\nu$ to columns. The scalar quantity $\overline{Z^{000}}=2\overline{\rho_\chi}$ vanishes for the total field, while several Lipkin densities, including off-diagonal ones, remain finite, revealing a purely anisotropic tensorial field structure.}
    \label{Zilch_numerical}
\end{figure*}
\begin{figure*}
    \centering
    \includegraphics[width=1\linewidth]{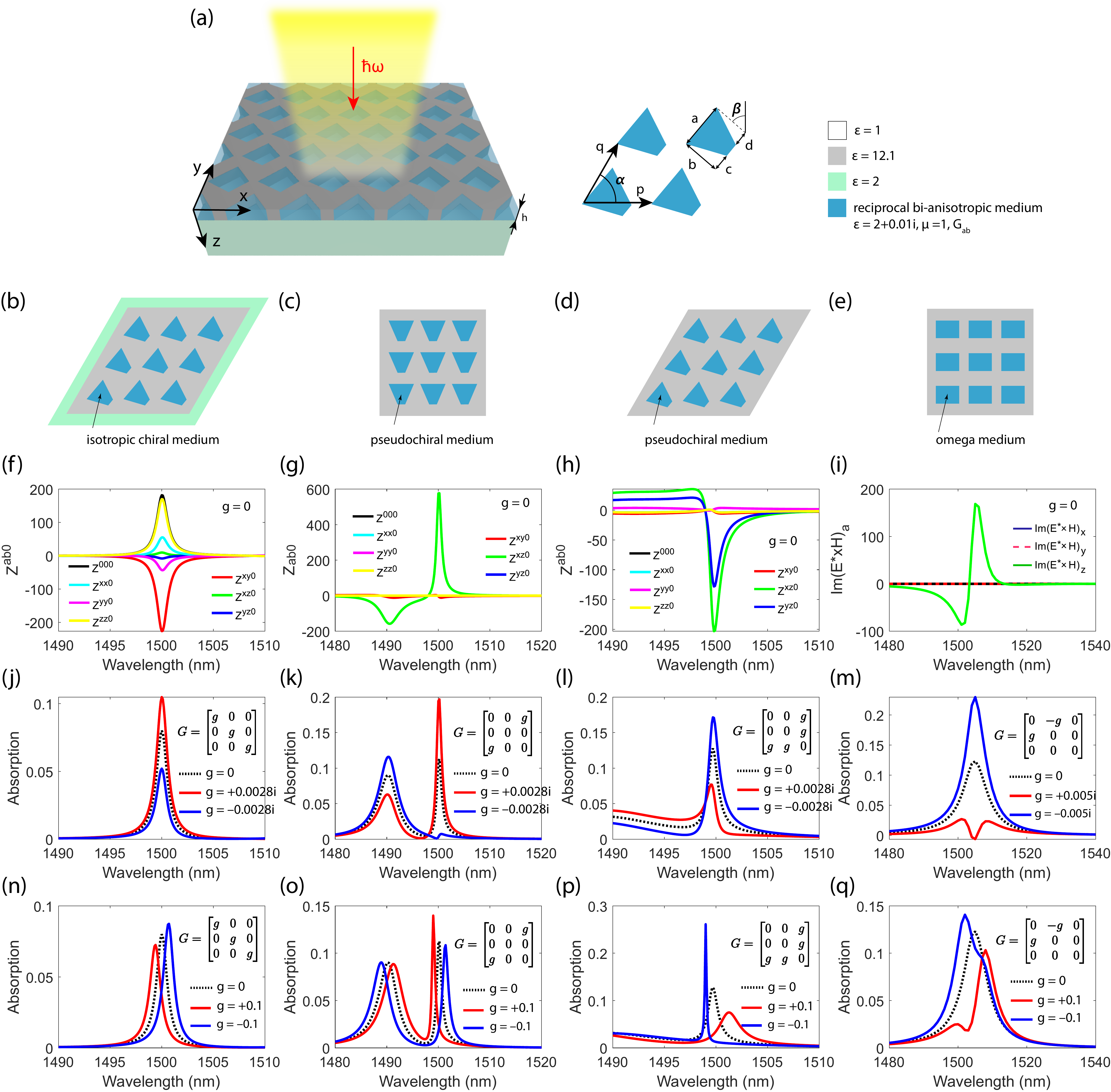}
        \caption{(a) Scheme of a metasurface containing rotated trapezoids. Colors indicate different materials. (b)--(e) Schematic diagrams of the metasurface geometries. (f)--(h) Time-averaged spatial-averaged densities $\overline{Z^{\mu\nu0}}$ for a p-polarized wave normally incident on the metasurface calculated for different geometries, indicated in panels (a)--(c). (i) Time-averaged spatial-averaged components of the reactive part of the local complex Poynting vector for a p-polarized wave normally incident light on the metasurface calculated for geometry, indicated in panel (e). Spatial averaging in panels (f)--(i) is performed over blue trapezoids. (j)--(q) Absorption spectra of the metasurface containing magneto-electric material calculated for cases with opposite $g$. Panels (j)--(m) corrrespond to real $g$, while panel (n)--(q) --- to imaginary $g$. Parameters of the metasurfaces are the following: panel (b) $p = 1020$~nm, $q = 860$~nm, $a = 0.25p$, $b = 0.34q$, $c = d = 0.48a$~nm, $\alpha = 100^\circ$, $\beta = 84^\circ$, $h = 166$~nm; panel (c): $p = 730$~nm, $q = 780$~nm, $a = 0.35p$, $b = 0.32q$, $c = d = 0.29a$~nm, $\alpha = 90^\circ$, $\beta = 0^\circ$, $h = 100$~nm; panel (d): $p = 1090$~nm, $q = 1080$~nm, $a = 0.47p$, $b = 0.46q$, $c = d = 0.48a$~nm, $\alpha = 66^\circ$, $\beta = 84^\circ$, $h = 70$~nm; panel (e) $p = 1170$~nm, $q = 770$~nm, $a = 0.30p$, $b = 0.32q$, $c = d = 0$, $\alpha = 90^\circ$, $\beta = 0^\circ$, $h = 192$~nm.}
    \label{fig:numerical_example2}
\end{figure*}



In the previous section, we showed that the scalar optical chirality density $\overline{\rho_\chi}=\overline{Z^{000}}/2$, despite its widespread use, fails to capture the full complexity of the interaction between bi-anisotropic media and low-symmetry fields. In particular, as evident from Table~\ref{tab:Gsym}, the parameter $\overline{\rho_\chi}$ is insensitive to pseudochiral materials. Below, we present numerical examples in which density components other than $\overline{Z^{000}}$ play a defining role. 

In the first example, we demonstrate that the electromagnetic field itself may have configurations in which the scalar optical chirality density vanishes while some Lipkin densities remain finite. To this end, we consider a simple yet instructive example of two monochromatic left-handed and right-handed \editS{waves} propagating in the $xz$-plane at angles of 0 and $\pi/4$ with respect to the $z$-axis (see Fig.~\ref{Zilch_numerical}). 

Figure~\ref{Zilch_numerical} shows the calculated density slices $\overline{Z^{\mu\nu0}}$ for each wave separately and for their superposition. One can see that although both waves have non-zero scalar optical chirality density ($\overline{Z^{000}}/2$), the corresponding parameter for the total field vanishes. This fact itself is not surprising, since it is intuitive that the superposition of two waves with opposite handednesses yields an achiral field. What is more intriguing is that, for the total field, the diagonal density components $\overline{Z^{110}}$ and $\overline{Z^{330}}$ are non-zero. From Table~\ref{tab:Gsym} and Eq.~\eqref{eq:Central_Tretyakov_Lipkin_result}, it follows that such a field, despite being achiral, is still sensitive to anisotropic chiral molecules \editS{with $G_{(11)}''\ne G_{(33)}''$.}

Note that optical chirality density $Z^{000}$/2, as well as the Lipkin densities $Z^{ab0}$, is quadratic in the electromagnetic field. Therefore, for a superposition of waves, 
the density magnitude of the total field is not generally equal to the sum of the corresponding densities of the individual waves. 

In the second example, we demonstrate different regimes of coupling between structured light and bi-anisotropic media. To this end, we consider a metasurface with trapezoidal inclusions (Fig.~\ref{fig:numerical_example2}). 
One can choose the geometry of the metasurface in such a way that it supports eigenmodes with various combinations of non-zero Lipkin densities or components of the reactive Poynting vector. The number of possible scenarios is very large; in this Article, we examine only four: one for inclusions filled with an isotropic chiral medium, two with a pseudochiral medium, and one with an omega medium. In all cases, we excite the electromagnetic field in the trapezoids using a normally incident p-polarized plane wave. The field distributions and absorption spectra are calculated using the Fourier modal method adapted for periodic slabs with bi-anisotropic materials \cite{smagin2026fourier}. In calculations, we employed 49 Fourier harmonics. For better convergence, factorization rules were used. 

\editS{Fig.~\ref{fig:numerical_example2}(b) displays a metasurface with rotated trapezoids on a substrate. In such a system, a normally incident $p$-polarized wave excites chiral electromagnetic field in trapezoids, as shown in Fig.~\ref{fig:numerical_example2}(f) by the time-average trapezoidal-average optical chirality densities calculated for the case of $G_{ab}=0$. Parameters of the metasurface are chosen in such a way that all density components are rather large in the vicinity of the resonant wavelength $\lambda\approx1497$~nm. We then introduce optical activity pseudotensor to the material of trapezoids $G_{ab}=g\delta_{ab}$ and calculate the absorption spectra for different values of the pseudoscalar $g$. Being chiral, such a field interacts differently with chiral bi-anisotropic materials [see distinct red and blue lines in Fig.~\ref{fig:numerical_example2}(f),(n)]. In particular, imaginary $g$ results in different absorption magnitudes for the opposite enantiomers, while real $g$ leads to the resonance frequency shift. 

More symmetric configurations are shown in Fig.~\ref{fig:numerical_example2}(c),(d), where only one or two independent Lipkin densities are sufficiently non-zero. In both cases, the non-zero density components are off-diagonal [Fig.~\ref{fig:numerical_example2}(g),(h)], and hence the fields in trapezoids are achiral. These fields  couple to pseudochiral material with corresponding non-zero components of optical activity pseudotensor [Fig.~\ref{fig:numerical_example2}(l),(m),(p),(q)]. Therefore, in this particular example, the absorption spectrum is sensitive only to the selected off-diagonal components of the material optical activity pseudotensor rather than to the entire pseudotensor $G_{ab}$ or to the scalar optical activity parameter $G$ used in \cite{Tang2010}. 

A similar strategy can be applied to couple the resonator field to an omega-medium, whose magnetoelectric response is described by a pseudotensor $G''_{ab}$ with off-diagonal components of opposite sign, $G''_{ab} = -G''_{ba}$ (see App.~\ref{App2}). According to Eq.~\eqref{eq:Central_Tretyakov_Lipkin_result}, such a medium does not couple to Lipkin density componets, but it can interact efficiently with fields that possess an enhanced non-zero reactive part of the Poynting vector, $\mathrm{Im}(\mathbf E^*\times \mathbf H)_a$. To achieve this, we again use the normally-incident $p$-polarized wave, which, as shown in Fig.~\ref{fig:numerical_example2}(i), excites a field in the metasurface  [Fig.~\ref{fig:numerical_example2}(e)] with a strongly enhanced component $\mathrm{Im}(\mathbf E^*\times \mathbf H)_3$. When an omega-medium with non-zero components $G''_{12} = -G''_{21}$ is introduced into this metasurface, the fourth term in Eq.~\eqref{eq:Central_Tretyakov_Lipkin_result} becomes active, leading to a pronounced difference in the absorption spectra for opposite orientation of the omega-medium, as confirmed by the calculated spectra shown in Fig.~\ref{fig:numerical_example2}(m),(q). 
}

\section{Conclusion}

In this work, we have revisited the Lipkin zilch tensor as a fundamental quantity characterizing electromagnetic chirality and provided a systematic physical interpretation of its components. \edit{In the normalization adopted here,} the scalar component \edit{satisfies $Z^{000}=2\rho_\chi$} and governs isotropic chiral \edit{light--matter interaction, whereas the Lipkin densities} $Z^{ab0}$ encode \edit{anisotropic tensorial} features of electromagnetic fields.

By \edit{extending the Tang--Cohen framework to an anisotropic electric-dipole--magnetic-dipole response}, we established that the \edit{symmetric part of the optical-activity pseudotensor is directly coupled to the Lipkin densities. This provided a clear physical in-
terpretation for the components of the zeroth slice of the
Lipkin zilch tensor. Additionally, we also have demonstrated that the antisymmetric part of optical-activity tensor is coupled to the reactive part of the Poynting vector.}

A simple numerical example illustrated that electromagnetic fields with vanishing scalar optical chirality may nevertheless possess a nontrivial anisotropic chiral structure, which remains invisible at the scalar level but is fully captured by the Lipkin densities. This observation highlights the limitations of scalar chirality measures and underscores the necessity of a tensorial description in structured optical fields.

The results presented here suggest that anisotropic electromagnetic chirality is a natural and physically relevant concept, particularly in systems where geometry and symmetry reduction play a central role. In this context, structured environments such as photonic crystals, metamaterials, and metasurfaces provide a promising platform in which specific \edit{Lipkin densities} can be selectively enhanced and probed. We anticipate that the tensorial framework developed in this work will prove useful for the systematic design and analysis of chiral \edit{light--matter} interactions in such geometrically engineered systems.

\section*{Acknowledgement}
This work was supported by the Russian Science Foundation (project 25-12-00454). The authors acknowledge Nikolay Gippius and Ilia Fradkin for a valuable discussion.

\refstepcounter{appnum}
\section*{Appendix~\theappnum}\label{App1}
\setcounter{equation}{0}
\renewcommand{\theequation}{A\arabic{equation}}

In this appendix we show that symmetric traceless $G_{(ab)}$ couples to $\overline{Z^{(ab)0}}$. We introduce the symmetric field bilinear
\begin{align}\label{eq:B_ab_definition}
    B_{ab}
    =
    \operatorname{Im}
    \left(
        E_a^*H_b
        -
        H_a^*E_b
    \right).
\end{align}
It is symmetric under $a\leftrightarrow b$. In terms of $B_{ab}$, the
time-averaged spatial Lipkin-density block can be written as
\begin{align}\label{eq:Zab0_B_relation}
    \overline{Z^{ab0}}
    =
    -
    \frac{\omega}{2}
    \left(
        \delta_{ab}B_{cc}
        -
        2B_{ab}
    \right).
\end{align}
Indeed, Eq.~\eqref{eq:Zab0_B_relation} reproduces the diagonal Lipkin
densities through the matrices $\mathcal G^{11}$, $\mathcal G^{22}$,
$\mathcal G^{33}$ from Eq.~\eqref{Gram matrices den 1}, and the off-diagonal densities through
$\mathcal G^{12}$, $\mathcal G^{13}$, $\mathcal G^{23}$ from Eq.~\eqref{Gram matrices den 2}. Taking the trace of Eq.~\eqref{eq:Zab0_B_relation} gives
\begin{align}
    \overline{Z^{110}}
    +
    \overline{Z^{220}}
    +
    \overline{Z^{330}}
    =
    -
    \frac{\omega}{2}B_{cc}
    =
    \overline{Z^{000}}.
\end{align}
Therefore the traceless Lipkin-density block is
\begin{align}\label{eq:traceless_Z_B_relation}
    \overline{Z^{(ab)0}} =
    \overline{Z^{ab0}}
    -
    \frac13
    \delta_{ab}
    \overline{Z^{000}} =
    \omega
    \left(
        B_{ab}
        -
        \frac13
        \delta_{ab}B_{cc}
    \right).
\end{align}
Since $G_{(ab)}\delta_{ab}=0$, contraction with $G_{(ab)}''$ removes the trace term in Eq.~\eqref{eq:traceless_Z_B_relation}. Hence
\begin{align}\label{eq:Gtraceless_Ztraceless_identity}
    \frac{\omega}{2}
    G_{(ab)}''B_{ab}
    =
    \frac12
    G_{(ab)}''
    \overline{Z^{(ab)0}}.
\end{align}
The pseudochiral contribution to the excitation rate is therefore
\begin{align}\label{eq:A_pseudochiral_sector}
    A^\pm_{\rm pseudo}
    =
    \pm
    \frac{\omega}{2}
    G_{(ab)}''
    \operatorname{Im}
    \left(
        E_a^*H_b
        -
        H_a^*E_b
    \right)
    =
    \pm
    \frac12
    G_{(ab)}''
    \overline{Z^{(ab)0}}.
\end{align}
Thus the symmetric traceless reciprocal response couples not to the optical chirality density, but to the traceless part of the Lipkin density block.

\refstepcounter{appnum}
\section*{Appendix~\theappnum}\label{App2}
\setcounter{equation}{0}
\renewcommand{\theequation}{B\arabic{equation}}

In this appendix we show that omega-type molecules can be described by $C_{nv}$, $n>2$ point groups. Pseudotensor $\mathbb{G}$ is invariant under transformation $R$ if
\begin{align}\label{app3:eq1}
    \mathbb{G} = \det(R) R\mathbb{G} R^T
\end{align}
where $R^T$ is transpose transformation matrix. First, consider the group $C_n$, where $n=1,2,3,...$ for now. Rotation around $z$ axis with $\varphi$ angle is
\begin{align}
    R_n =
    \begin{pmatrix}
        r & 0 \\
        0 & 1
    \end{pmatrix},
    \qquad
    r = 
    \begin{pmatrix}
        c & -s \\
        s & c
    \end{pmatrix}.
\end{align}
Here $c=cos\varphi$, $s=sin\varphi$ and $\varphi=\frac{2\pi}{n}$. For this rotation $\det(R_n)=+1$. Consider the pseudotensor $\mathbb{G}$ as follows
\begin{align}
\begin{split}
    \mathbb{G}  &=
    \begin{pmatrix}
        M & u \\
        v^T & G_{zz}
    \end{pmatrix},
    \\
    M = 
    \begin{pmatrix}
        G_{xx} & G_{xy} \\
        G_{yx} & G_{yy}
    \end{pmatrix}, 
    \quad
    u &= 
    \begin{pmatrix}
        G_{xz} \\ G_{yz}
    \end{pmatrix},
    \quad
    v = 
    \begin{pmatrix}
        G_{zx} \\ G_{zy}
    \end{pmatrix}.
\end{split}
\end{align}
Then, we have four symmetry conditions from Eq.~\eqref{app3:eq1}
\begin{align}\label{app3:eq4}
\begin{split}
    &M = rMr^T,\\
    &u = ru, \\
    &v = rv, \\
    &G_{zz} = G_{zz}
\end{split}
\end{align}
The last expression in Eq.~\eqref{app3:eq4} does not impose any restrictions on $G_{zz}$. The determinant of the second system in Eq.~\eqref{app3:eq4} equals to $2(1-c)$ and therefore for nontrivial rotation $\varphi\neq0$ and for $n>1$ this system has only trivial solution $G_{xz} = G_{yz} = 0$. From the same thoughts for $v$ we conclude that $G_{zx} = G_{zy} = 0$. Next, we can rewrite the first system in Eq.~\eqref{app3:eq4} as $rM = Mr$ because $rr^T=1$. From this expression we conclude that $s(G_{xy} + G_{yx}) = 0$. For $n>2$ $s\neq0$ and then $G_{xy} = -G_{yx}$. Moreover, for $n>2$ it appears that $G_{xx} = G_{yy}$.

After all, for the $C_{n}$, $n>2$ point groups the pseudotensor $\mathbb{G}$ must have the form
\begin{align}
    \mathbb{G} =
    \begin{pmatrix}
        A & B & 0 \\
        -B & A & 0 \\
        0 & 0 & C
    \end{pmatrix}.
\end{align}
Finally, we impose a condition of vertical mirror. As an example we consider the mirror in $xz$ plane with reflection operation
\begin{align}
    R_{xz} =
    \begin{pmatrix}
        1 & 0 & 0 \\
        0 & -1 & 0 \\
        0 & 0 & 1
    \end{pmatrix}.
\end{align}
From Eq.~\eqref{app3:eq1} we have
\begin{align}
    \mathbb{G} = - R_{xz} \mathbb{G} R^T_{xz}.
\end{align}
This expression vanishes the diagonal components in the pseudotensor $\mathbb{G}$, whereas nondiagonal components do not change. At the end for $C_{nv}$, $n>2$ point groups we have the following pseudotensor $\mathbb{G}$
\begin{align}
    \mathbb{G}_\Omega = 
    \begin{pmatrix}
        0 & g & 0 \\
        -g & 0 & 0 \\
        0 & 0 & 0
    \end{pmatrix}.
\end{align}

\bibliography{ref}

\bibliographystyle{unsrt}

\end{document}